\documentclass[english]{achemso}
\usepackage[T1]{fontenc}
\usepackage{amstext}
\usepackage{graphicx}
\usepackage{subscript}

\makeatletter

\title{Automated Discovery of Wurtzite Solid Solutions with Enhanced Piezoelectric Response}

\author{Drew Behrendt}
\affiliation{Department of Chemistry, University of Pennsylvania, Philadelphia, Pennsylvania, 19104-6323, USA}
\author{Sayan Banerjee}
\affiliation{Department of Chemistry, University of Pennsylvania, Philadelphia, Pennsylvania, 19104-6323, USA}
\author{Jiahao Zhang}
\affiliation{Department of Chemistry, University of Pennsylvania, Philadelphia, Pennsylvania, 19104-6323, USA}
\author{Andrew M. Rappe}
\affiliation{Department of Chemistry, University of Pennsylvania, Philadelphia, Pennsylvania, 19104-6323, USA}
\email{rappe@sas.upenn.edu}

\makeatother

\usepackage{babel}
\begin{document}

\section*{Abstract:}

While many piezoelectric materials are known, there is still great potential to improve on the figures of merit of existing materials through compositional doping, forming solid solutions. Specifically, it has been shown that doping and alloying wurtzite-structured materials can improve the piezoelectric response; however, a vast compositional space has remained unexplored. In this work, we apply a multi-level screening protocol combining machine learning, chemical intuition, and thermodynamics to systematically discover dopant combinations in wurtzite material space that improve the desired piezoelectric response. Through our protocol, we use computationally inexpensive screening calculations to consider more than 3000 possible ternary wurtzite solid solutions from 9 different wurtzite base systems: AlN, BeO, CdS, CdSe, GaN, ZnO, ZnS, ZnSe, and AgI. Finally, based on thermodynamic analysis and explicit piezoelectric response calculations, we predict 11 materials with improved piezoelectric response, due to the incorporation of electropositive dopants.

\section*{Main Text:} 

Piezoelectrics are non-centrosymmetric materials that are capable of interconverting mechanical and electrical energy for a variety of applications.\cite{Nurettin} Piezoelectrics provide the basis for microelectronic energy harvesting, acoustic wave devices, actuators, and other devices that are widely used in research, industry, and military applications. \cite{Song2021,Yoshioka2021,piazza_12} The energy efficiency and power output of these materials increases with increasing relative piezoelectric response, but decreases with increasing dielectric constant. Perovskite ferroelectrics are prototypical piezoelectrics; however, while perovskites have a very high piezoelectric response, they also possess high dielectric constants.\cite{POPLAVKO2020161,piazza_12} Furthermore, these materials often lose their beneficial properties at high temperatures.\cite{POPLAVKO2020161} An alternate material class is wurtzite; though wurtzites have smaller piezoelectric constants than their perovskite counterparts, they are known for their low dielectric constants and high-temperature performance. \cite{Ambacher2021,Momida2016,Momida2018} Furthermore, these materials are highly compatible with complementary metal-oxide-semiconductor (CMOS) technology, which makes them highly attractive for use in piezoelectric applications. \cite{Song2021,Yoshioka2021} Doping has long been used to improve the electronic properties of host materials; for example, perovskite lead magnesium niobate (PMN) is alloyed with lead titanate (PT) to enhance the temperature stability and magnitude of piezoelectricity. Adding elements into common wurtzite materials, such as doping aluminum nitride with scandium or zinc oxide with magnesium, has been shown to appreciably increase the piezoelectric response and even induce ferroelectricity, which is of interest for many additional applications.\cite{Momida2016,Noor-A-Alam2019,Py-Renaudie2020,Zywitzki2017,Krishnamoorthy2021,Moriwake2020} 


With the recent growth of machine learning (ML) applications in materials science, many methods have been proposed to speed up materials discovery using ML-based methods.\cite{sanchez2018inverse,tao2021nanoparticle,ulissi2017address,zhong2020accelerated,raccuglia2016machine,zunger2018inverse,butler2018machine,ren2022invertible,xin2022catalyst} Feature selection, where one finds which input features are most correlated to a target output, is a primary challenge of current ML applications in big data. Choosing features is at the heart of this challenge, since it is inherently subjective, and the only way to find meaningful correlations is to have representative and meaningful features. Once chosen and selected, however, features can be particularly useful in rational materials design because of their role in providing an interpretative picture of the underlying physics. Feature selection approaches have been shown to be successful for fields spanning from materials discovery\cite{raccuglia2016machine,balachandran2018importance} to chemical reaction development using homogeneous\cite{milo2014interrogating,singh2020unified,banerjee2018machine,dos2021navigating} and heterogeneous\cite{esterhuizen2022interpretable,Wexler2018,esterhuizen2020theory,xin2022catalyst,gao2022breaking} catalysts. Many recent works have identified a  "material gene" to perform advanced analysis of material structure-property relations, where the material gene corresponds to the most important feature(s) that correlate with and therefore influencing a target property.\cite{Wexler2018,Mazheika2022,Balachandran2011,He2021,Lyu2021,Meredig2014,Ward2016,Xu2021,Behrendt} The features can be broadly classified into two different types: primitive and calculated. Primitive features, or primitives, are purely based on summary statistics of atomic data such mass, location on the periodic table, electronegativity, and charges.\cite{Ward2016} Calculated features, reffered to as proxies, involve calculations such as \emph{ab initio} density functional theory calculations, and these include ionic charges, bond lengths, and bond angles. It is worth noting that, to be most useful, calculated features of a material should be simpler and less time consuming to calculate than the target property of interest. For example, calculation of the lattice parameters requires much less computational time than piezoelectric tensor calculations. Our multi-level screening protocol is based on the combination of these two types of featurization.

In the current work, we have investigated nine different base wurtzite materials and their solid solutions with every relevant metal and metalloid element, over 3000 possible candidate materials. We narrow down this space to 30 solid solutions which are predicted to improve the piezoelectric response. Finally, we perform thermodynamic analysis to predict 11 best candidates for future experimental verification. An overview of this workflow is shown in Fig 1. The first level of screening is designed to find one screening proxy to use in place of expensive piezoelectric calculations. The second level of the screening serves to reduce the space of all examined ternary solid solutions to just a few suggested high-value materials using automated machine learning candidate selection and cost of the suggested dopants. 

\begin{figure}
    \centering
    \includegraphics[width=1\linewidth]{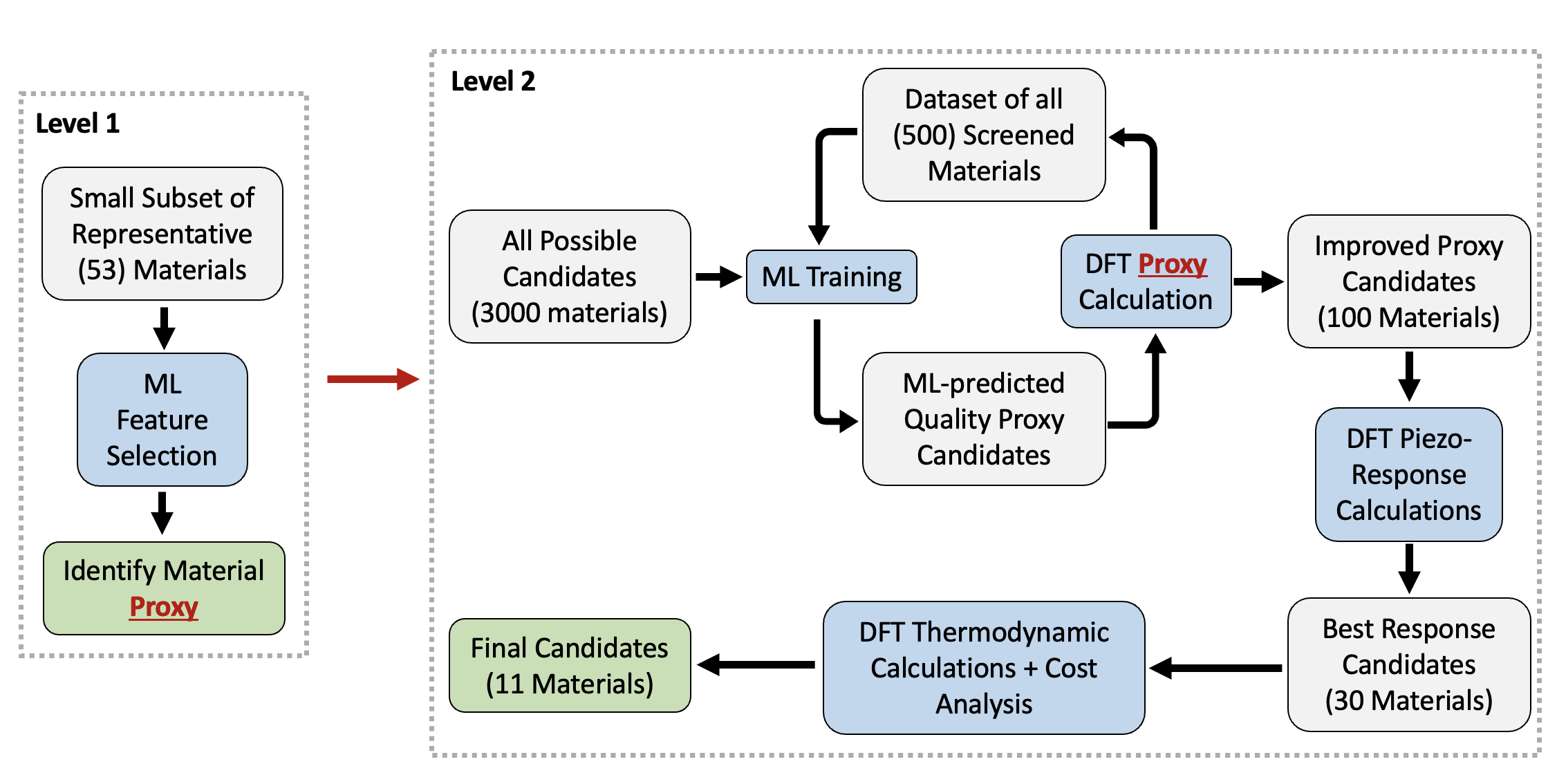}
    \caption{Multi-level screening workflow to explore the compositional phase space of the nine different wurtzite base materials to maximize the piezoelectric tensor component $e_{33}$. The grey, blue, and green boxes represent data sets, methods, and deliverables, respectively.}
    \label{fig:workflow}
\end{figure}

We consider a 2 $\times$ 2 $\times$ 1 wurtzite supercell consisting of 8 metal and 8 non-metal sites. For doped materials, 2 of the 8 cation sites have been replaced. The relative positions of the dopants within the unit cell has been shown to affect the piezoelectric response, \cite{Momida2016} so it is kept constant throughout this entire study to compare elemental doping effects on an equal footing. The calculations for the $e_{33}$ component of the piezoelectric tensor are performed using the Quantum Espresso \cite{Giannozzi_2009} software package with optimized, norm-conserving pseudopoentials generated by OPIUM. \cite{Rappe90,Rappe99} To do this, the system is strained along the $c$ axis for five different values ($-$1.0 to +1.0 percent strain), and then atomic positions are allowed to relax. The polarization is then calculated using the Berry's phase method.\cite{vanderbilt2000berry,vanderbilt2018berry} The slope of polarization \emph{vs.} strain was calculated and taken to be the $e_{33}$ component\cite{saghi1998first} of the piezoelectric tensor for the calculated material. For doping, all combinations of metallic elements that would preserve charge neutrality are considered for each system. For example, combinations of (+3,+3) or (+2,+4) dopants are considered for AlN, whereas for ZnO (+1,+3) or (+2,+2) are screened. Five different machine learning (ML) methods, linear regression, least absolute shrinkage and selection operator (LASSO), ridge, recursive feature elimination (RFE), and random forest (RF), from python's sklearn are employed.\cite{scikit-learn} By training multiple machine learning algorithms on each round of the active learning, we can ensure faster sampling of a larger candidate space since the algorithms will select different types of materials from the limited initial database, and ultimately we can ensure that if any algorithm finds a candidate to be viable then it is screened. Additionally, after training these different methods we can compare the relative effectiveness of each in predicting the material proxy from the primary features.

To identify the best screening proxy, we started with a moderate dataset of additive elements in the most commonly used wurtzite, aluminium nitride (AlN), to improve the piezoelectric response (Level 1 in Fig.\ \ref{fig:workflow}). There have been multiple studies showing the effects of co-doping into AlN. Various material descriptors, especiallly the lattice $\frac{c}{a}$ ratio, have been identified as key properties that correlate with the piezoelectric response. \cite{Noor-A-Alam2021,Momida2018,Momida2016,Iwazaki2015,Py-Renaudie2020,Tholander2015} In total, 53 materials (see supplementary materials) are included in our initial dataset, and the results for the piezoelectric response are summarized in Table \ref{tab:my_label}. Notably, AlN doped with boron and scandium together shows a slight increase in the piezoelectric constant relative to AlN with scandium alone, which is currently attracting great interest.\cite{Song2021,Yoshioka2021}

\begin{center}
    \begin{table}[]
        \centering
        \begin{tabular}
        {|c|c|}
        \hline
            Dopants & $e_{33}$ \\
            \hline
            None & 1.46 \\
            Sc & 1.86 \\
            Sc,B & 1.90 \\
            Mg,Hf & 1.84 \\
            Mg,Ti & 1.60 \\
            \hline
        \end{tabular}
        \caption{Notable piezoelectric response for dopants in AlN, from among the initial dataset of 53 materials. }
        \label{tab:my_label}
    \end{table}
\end{center} 

\begin{figure}
    \centering
    \includegraphics[width=1\linewidth]{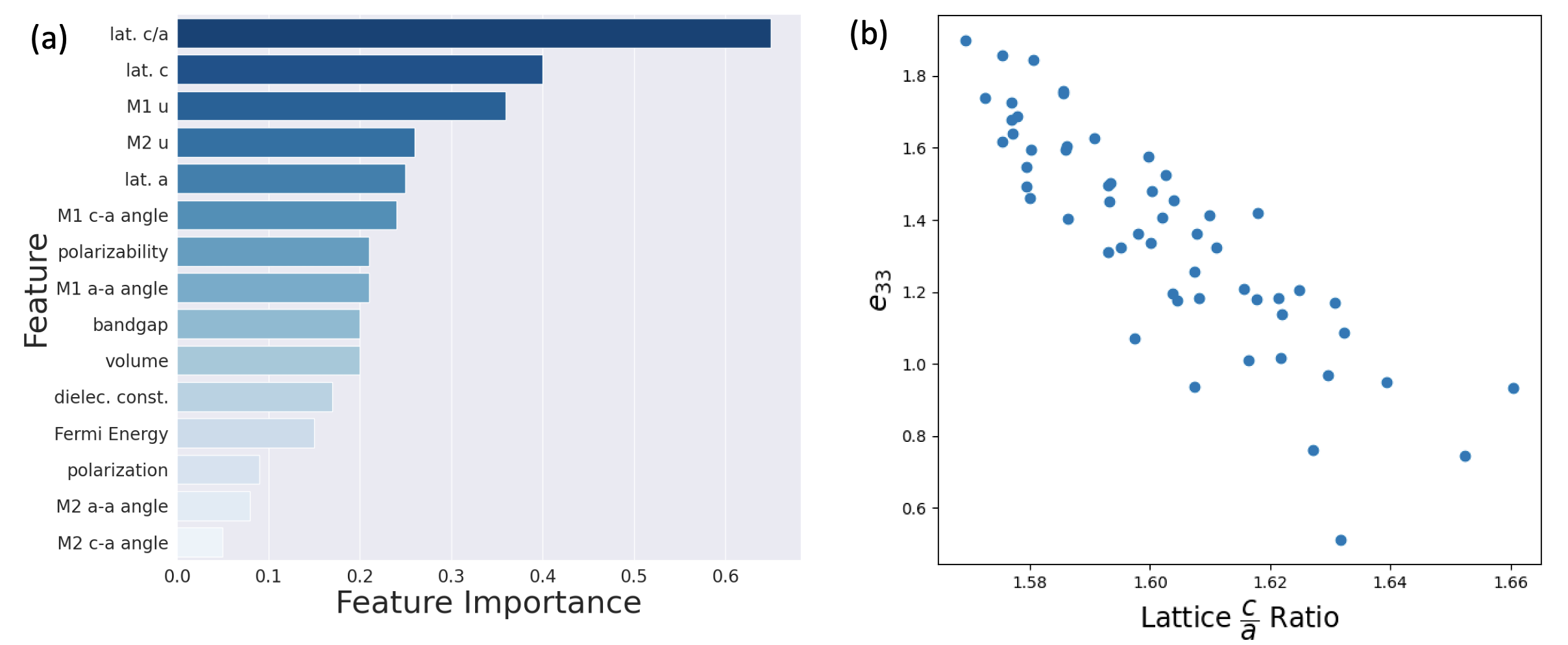}
    \caption{Feature selection in the initial dataset of 53 co-doped AlN materials: (a) ML feature selection to identify the most important material proxy. (b) Scatter plot showing the correlation between lattice $\frac{c}{a}$ and $e_{33}$ response. Abbreviated proxies in (a) are shown in full in the supplementary material. All of the important proxies represent optimized bond lengths ($u$ parameter), angles, lattice parameters, or simple electronic properties.}
    \label{fig:proxy}
\end{figure}

For each material in the dataset, we used 15 possible candidate proxy properties based on previous work on material genes and wurtzite characterization (supplementary material). We then used machine learning feature selection to find the material proxy that is, on average, most correlated to $e_{33}$, and the results are shown in Figure 2. Note that the feature importance for each feature is averaged across all of the ML methods for proxy selection. We find the lattice $\frac{c}{a}$ ratio to be the most important feature (Fig.~\ref{fig:proxy}a), which has also been reported in previous literature for host materials alone and based on manual manipulation of the lattice parameters for sample systems.\cite{Momida2018,Noor-A-Alam2019,Py-Renaudie2020}. 
Therefore, a lower lattice $\frac{c}{a}$ ratio for wurtzite solid solutions corresponds to a higher value of $e_{33}$ (Fig.~\ref{fig:proxy}b), in accordance with previous works.\cite{Momida2018,Noor-A-Alam2019,Py-Renaudie2020} We use the $\frac{c}{a}$ ratio as a screening proxy for our automated workflow (Fig.\ \ref{fig:proxy}b).

Based on the lattice $\frac{c}{a}$ ratio proxy, we generated a set of over 3000 possible candidate solid solutions from 9 different wurtzite base systems: AlN, BeO, CdS, CdSe, GaN, ZnO, ZnS, ZnSe, and AgI. Starting from the initial 53-material dataset and base materials, we then used machine learning predictions to iteratively select candidates from the 3000 possible combinations to screen. For each atom, uncalculated descriptors were chosen to reflect the fundamental chemical characteristics including: atomic number, atomic mass, elemental melting temperature, charge, column and row on periodic table, atomic radius \cite{Shannon1976}, electronegativity\cite{Tantardini2021}, and preferred valence. Primitive features were collected by taking the average, standard deviation, minimum, maximum, and range of the atomic descriptors for each material (a total of 50 primitives for each material). The key screening steps are: i) training a ML model using the primitives of the materials in the database in a specific iteration, ii) using the trained ML model to predict which of the unstudied candidates will have the lowest $\frac{c}{a}$ ratio to screen next, and iii) adding the newly screened materials to the database (the iterative ML screening in Fig. \ref{fig:workflow}). This process was repeated until the best 500 candidates were screened. 

During the protocol, it became clear that doped versions of certain base materials are much more likely to have a lower lattice ratio than others (specifically for BeO, AlN, ZnO, GaN). Therefore, in order to screen the best candidates for each base material, once solid solutions of a given base material are no longer predicted to give significant improvement to the proxy of the host, the rest of the elemental combinations of that base material were removed from the remaining active learning candidates. This ensures that we suggest new variants for all nine different base materials, even if some have a higher propensity toward high $\frac{c}{a}$ ratio than others. Additionally, some elements are inherently unstable in the host material and the relaxation calculations fail for a variety of reasons. In this circumstance, a high value of $\frac{c}{a}$ ratio of 2.0 was assigned to these co-dopants so that the ML predictions would avoid similar materials. 

\begin{figure}
    \centering
    \includegraphics[width=1.0\linewidth]{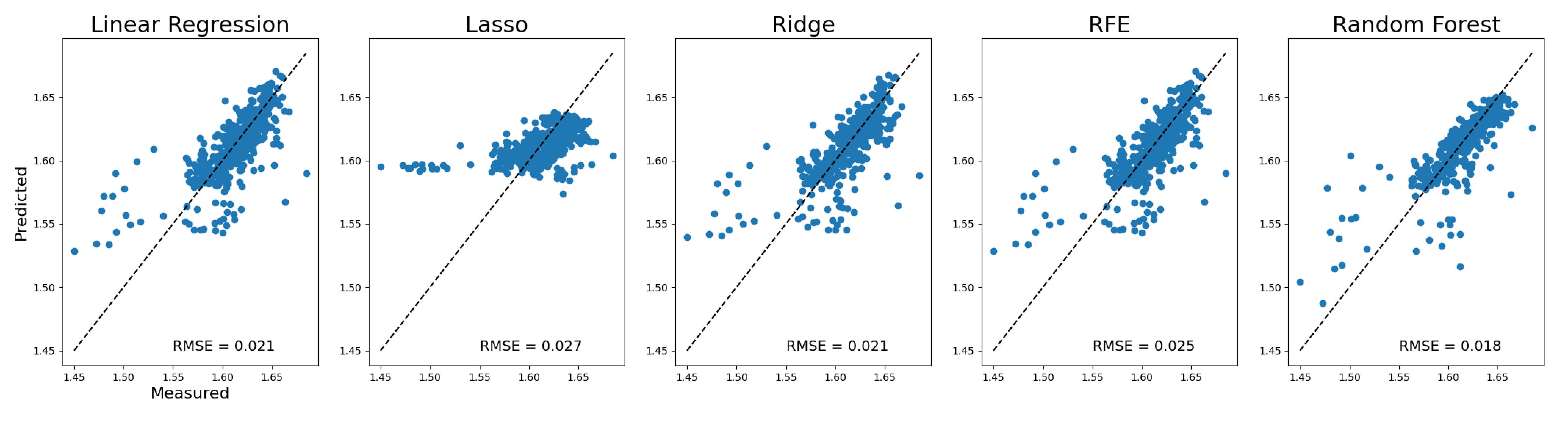}
    \caption{Validation of the predictive power of ML models. The low root mean square error (RMSE) for multiple models indicates that all the viable candidates are screened.}
    \label{fig:cross-validation}
\end{figure}

After screening, we performed five-fold cross validation for each ML model. This is done to ensure that the ML models are accurately predicting the lattice ratios from the primitive features so that we can be confident that all viable candidates are screened correctly; the results are shown in Fig. \ref{fig:cross-validation}. We find that the random forest algorithm worked best to predict the lattice ratios. However, with the exception of the LASSO regression, all of the methods used are relatively accurate at predicting the proxy from primitives, which is evident from the low RMSE given in Fig. \ref{fig:cross-validation}.  All of the methods have particular difficulty with materials that are predicted to have $\frac{c}{a}$ ratios near the average yet are calculated by DFT to possess extremely low lattice ratios; this is in part due to the fact that most of these extreme values represent materials that become unstable and deviate from the original wurtzite structure. Furthermore, because the only poor predictions are in the low $\frac{c}{a}$ regime, we conclude that the unscreened materials predicted to have high $\frac{c}{a}$ are unlikely to be good candidate piezoelectrics. At the end of the screening, only about 1/5 of all the screened solid solutions are found to be insulating and actually effective at reducing the lattice ratio for the parent material; these materials are then verified by subsequent $e_{33}$ calculations in accordance with the workflow in Fig. \ref{fig:workflow}. 

\begin{figure}
    \centering
    \includegraphics[width=0.7\linewidth]{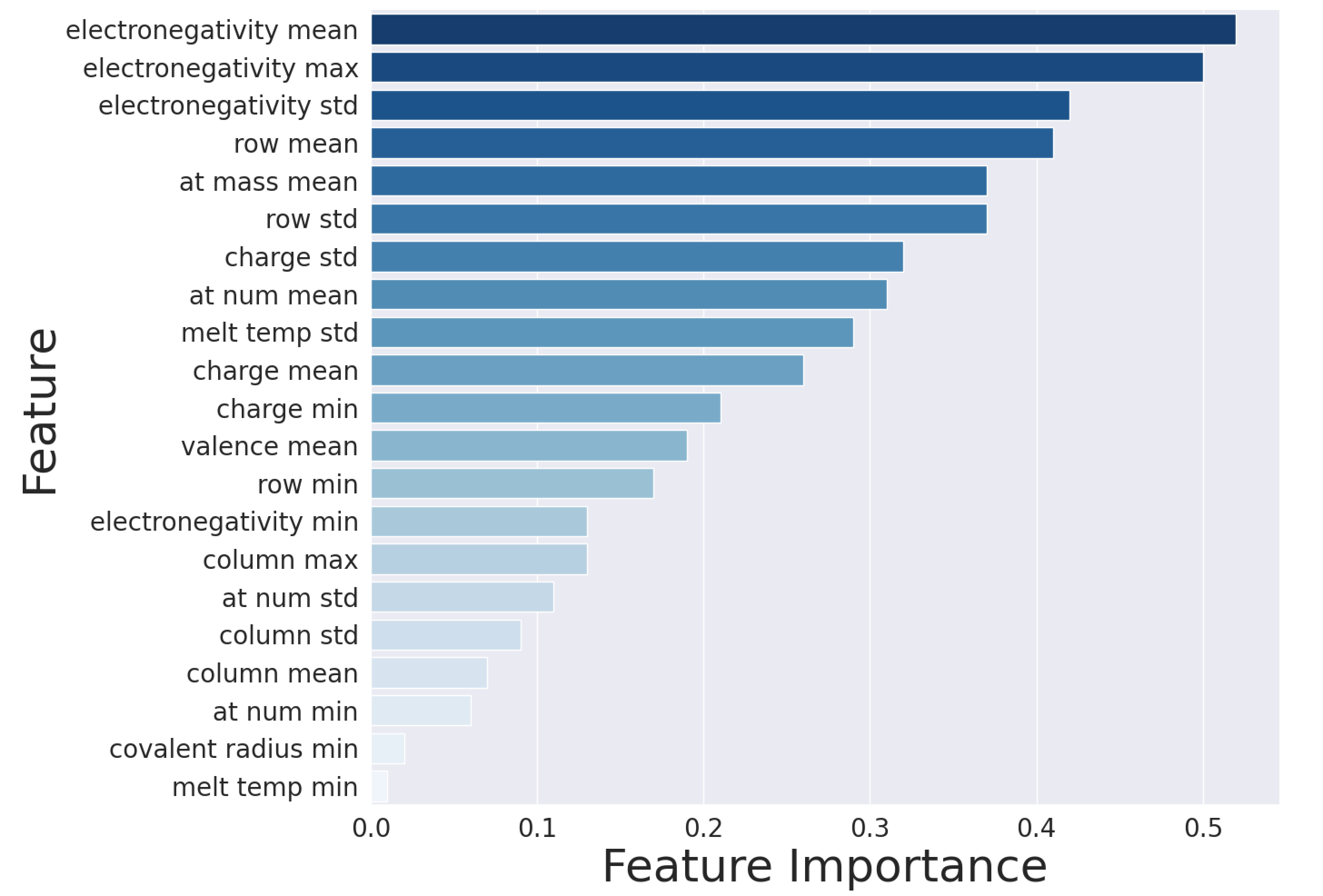}
    \caption{Importance of primary features to predicting lattice $\frac{c}{a}$ ratio. "At" stands for atomic, "std" for standard deviation, and the melting temperature is of the material for each element in solid form. All features involved no calculations are are summary statistics for the constituent elements in each candidate material.}
    \label{fig:primselect}
\end{figure}

As shown in Fig. \ref{fig:cross-validation}, the LASSO algorithm had the worst fitting for the ML algorithms. By design, LASSO tends to lead to only a few features carrying very high weight in the trained model; thus, the poorer fitting indicates that many primitive features are needed in combination to accurately predict the lattice $\frac{c}{a}$ ratio. However, determination of the most important features still can provide insight to the structure-property relations, guiding which elements are best to add to each wurtzite system. Since most of the ML algorithms trained on the primitive material features provide reliable prediction of the resulting $\frac{c}{a}$ ratio and insight into the possible piezoelectric response, we performed feature selection to see which primaries were most important (Fig. \ref{fig:primselect}). Two major atomic characteristics stand out as essential to a low $\frac{c}{a}$ ratio and consequently improved piezoelectric response:  electronegativity and mass. Materials containing extremely highly electronegative elements as the anion, oxygen and nitrogen in particular, had characteristically lower $\frac{c}{a}$ ratios. Additionally, materials with a high standard deviation of electronegativity in constituent elements, containing extremely electropositive elements as well, tended to have lower $\frac{c}{a}$ ratios. With the high importance of mean row and atomic mass, materials with a low average row and atomic mass also tended to have low $\frac{c}{a}$ ratios. This means that the addition of small, electropositive elements to AlN, ZnO, and BeO will reliably lead to heightened electronic response in wurtzites. This finding is aligned with recent evidence of Sc and B enhancing piezoelectricity in AlN, Mg in ZnO, and even the classic example of adding Mg and Nb to lead titanate. However, since the LASSO algorithm had relatively poor predictive power, we posit that these primitives alone are not enough to predict the $\frac{c}{a}$ ratio, and that all (or almost all) of the primitives provide enough information for accurate prediction of the lattice ratio and piezoelectric properties from elemental descriptors. 

\begin{figure}
    \centering
    \includegraphics[width=0.7\linewidth]{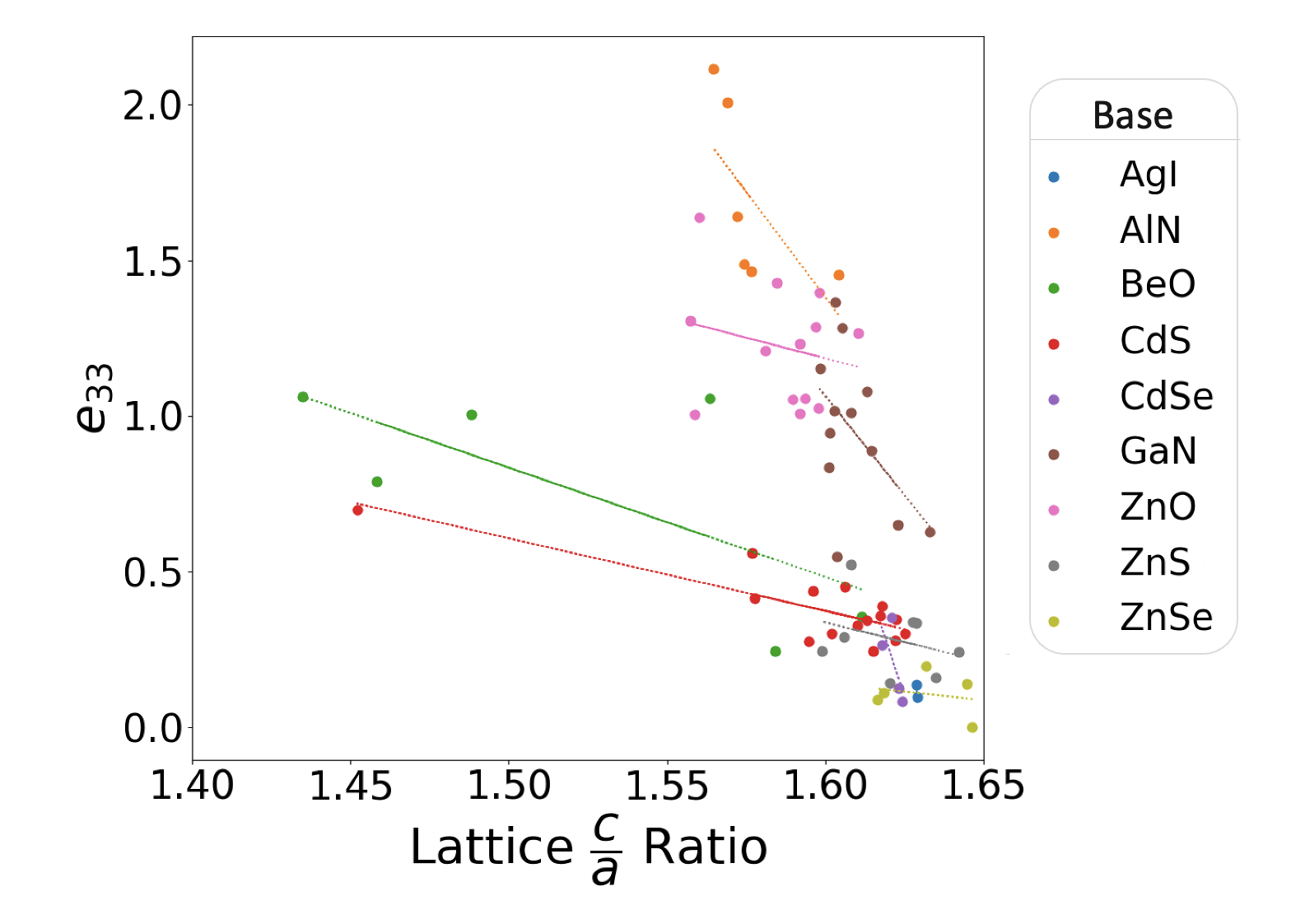}
    \caption{Lattice ratios and piezoelectric responses for the final screened materials. Each color corresponds to a base material and its respective co-dopants.}
    \label{fig:selected}
\end{figure}

During the verification of $e_{33}$ using Berry's phase polarization calculations (DFT response calculations in Fig. \ref{fig:workflow}), we find that while all good piezoelectrics have low $\frac{c}{a}$, not all wurtzite solid solutions with low $\frac{c}{a}$ are good piezoelectrics. For example, some of the the greatly reduced $\frac{c}{a}$ ratio systems unphysically distort the base material out of the wurtzite phase, leading to an unstable system. We observe that BeO in particular often becomes unstable with the addition of other elements. Furthermore, we find that the trend of lower $\frac{c}{a}$ ratio correlating with higher piezoelectric constant is not a single linear trend for all different wurtzite materials, as suggested in previous work,\cite{Momida2018} but instead we find a linear trend for dopants within each given base material, as shown in Fig. \ref{fig:selected}. The resulting 30 materials from this final screening that notably increased the piezoelectric coefficient of their respective base material are listed in Table 2. All of these materials are predicted to increase the $e_{33}$ of the parent material. However, not all are good candidates for functional materials due to thermodynamic and practical concerns (Fig. \ref{fig:workflow}). For each material, the best candidate doping combinations after considering such factors are highlighted in bold in Table \ref{tab:best}. 

While the proxy $\frac{c}{a}$ ratio helps to drastically reduce the computational expense for the materials screening, it is important to further validate ML-based predictions. The low $R^2$ values in Fig.~\ref{fig:selected} indicate that there are complicating factors beyond the $\frac{c}{a}$ ratio that govern the piezoelectric properties, particularly for zinc-containing materials (Fig.~\ref{fig:selected}). Additional factors, such as system Born effective charges, are discussed in the supplementary material and will be investigated in the future. However, we report that as a screening parameter, the $\frac{c}{a}$ proxy has been effective to find valuable candidate wurtzite solid solutions which can improve the piezoelectric response for all nine base materials. 

\begin{center}
    \begin{table}[]
        \centering
        \begin{tabular}
        {|c|c|c|c|c|c|}
        \hline
            Dopants & c/a & e$_{33}$ & Improvement & Formation Energy & Avg. Dopant Cost\\
            (Base) &  & (C/m$^2$) &  & (eV) & (USD/kg) \\
            \hline
            (AlN) & 1.604 & 1.46 & - & - & - \\
            B,Y & 1.572 & 1.64 & 1.12x & 4.299 & 1,200\\
            Mg,Hf & 1.581 & 1.84 & 1.26x & 2.129 & 700\\
            Sc & 1.575 & 1.86 & 1.27x & -0.069 & 15,000\\
            \textbf{Sc,B} & \textbf{1.569} & \textbf{1.90} & \textbf{1.30x} & \textbf{2.813} & \textbf{8,700}\\
            Be,Zr & 1.569 & 2.01 & 1.37x & 3.232 & 860\\
            \textbf{Be,Hf} & \textbf{1.565} & \textbf{2.12} & \textbf{1.45x} & \textbf{2.937} & \textbf{1,100}\\
            \hline
            (BeO) & 1.612 & 0.356 & - & - & - \\
            Li,Y & 1.458 & 0.791 & 2.22x & 2.338 & 75\\
            Li,Sc & 1.376 & 0.954 & 2.68x & 2.955 & 7,560\\
            Ga,Na & 1.488 & 1.01 & 2.84x & 9.267 & 140\\
            \textbf{Mg} & \textbf{1.435} & \textbf{1.06} & \textbf{2.98x} & \textbf{4.211} & \textbf{2}\\
            \textbf{Al,Li} & \textbf{1.563} & \textbf{1.06} & \textbf{2.98x} & \textbf{3.593} & \textbf{60}\\
            \hline
            (CdS) & 1.625 & 0.302 & - & - & - \\
            Be,Sr & 1.577 & 0.561 & 1.86x & -1.161 & 420\\
            \textbf{B,K} & \textbf{1.452} & \textbf{0.699} & \textbf{2.22x} & \textbf{1.811} & \textbf{1,200}\\
            \hline
            (CdSe) & 1.624 & 0.082 & - & - & - \\
            Na,Sc & 1.618 & 0.264 & 3.22x & -2.627 & 7,500\\
            \textbf{Ca,Mg} & \textbf{1.621} & \textbf{0.353} & \textbf{4.30x} & \textbf{-3.764} & \textbf{5}\\
            \hline
            (GaN) & 1.633 & 0.629 & - & - & -\\
            Mg,Ti & 1.614 & 0.889 & 1.41x & -2.628 & 3\\
            Be,Zr & 1.601 & 0.946 & 1.50x & -1.750 & 850\\
            B,Sc & 1.608 & 1.01 & 1.61x & -2.029 & 8,700\\
            Sc & 1.613 & 1.08 & 1.72x & -4.935 & 15,000\\
            \textbf{B,Y} & \textbf{1.598} & \textbf{1.15} & \textbf{1.83x} & \textbf{-0.886} & \textbf{1,200}\\
            \hline
            (ZnO) & 1.610 & 1.27 & - & - & - \\
            B,Na & 1.557 & 1.31 & 1.03x & -0.346 & 1,200\\
            Mg & 1.598 & 1.40 & 1.10x & -4.369 & 3\\
            Be,Mg & 1.585 & 1.43 & 1.13x & -3.750 & 420\\
            \textbf{Be,Ca} & \textbf{1.560} & \textbf{1.64} & \textbf{1.29x} & \textbf{-3.555} & \textbf{420}\\
            \hline
            (ZnS) & 1.642 & 0.243 & - & - & -  \\
            Be,Cd & 1.629 & 0.335 & 1.38x & 1.452 & 420\\
            Be,Mg & 1.627 & 0.340 & 1.40x & -0.276 & 420\\
            \textbf{Be,Ca} & \textbf{1.608} & \textbf{0.522} & \textbf{2.15x} & \textbf{-0.620} & \textbf{420}\\
            \hline
            (ZnSe) & 1.647 & 0.001 & - & - & -  \\
            \textbf{Na,Sc} & \textbf{1.644} & \textbf{0.140} & \textbf{140x} & \textbf{-1.896} & \textbf{7500}\\
            Be,Mg & 1.632 & 0.198 & 198x & 0.154 & 420\\
            \hline
            (AgI) & 1.629 & 0.098 & - & - & -  \\
            \textbf{Li} & \textbf{1.628} & \textbf{0.138} & \textbf{1.41x} & \textbf{-4.948} & \textbf{120}\\
            \hline
        \end{tabular}
        \caption{Notable piezoelectric response for co-doped systems. Base materials are in parenthese and materials written in bold are of particular interest for increasing the piezoelectric constant of the base system.}
        \label{tab:best}
    \end{table}
\end{center} 

Overall, through a multi-step screening protocol and high-throughput study, we identify previously unstudied solid solution candidates to improve the piezoelectric response of nine chosen wurtzite base materials. During this process, we emphasize the fundamental relation of $\frac{c}{a}$ lattice ratio in wurtzites to its respective piezoelectric response and use it as a proxy to develop a computationally inexpensive protocol for piezoelectric material discovery. Furthermore, we are able to support the idea that primitives can be effectively used in machine learning methods to predict basic material properties such as lattice parameters, given a sizeable dataset. Finally, we propose the best set of candidates for further experimental verification. We hope that the present work serves as a practical example in the design of materials with improved material properties through computationally efficient multi-step high-throughput studies.

\clearpage{}

\section*{Corresponding Author:}

{*}E-mail: rappe@sas.upenn.edu

\section*{ORCID:}

Drew Behrendt: 0000-0003-4701-2722; Sayan Banerjee: 0000-0002-8586-9236; Jiahao Zhang:0000-0002-8284-8122; Andrew M. Rappe: 0000-0003-4620-6496

\section*{Acknowledgements:}
D.B. and J.Z. thank the U. S. Department of Energy, Office of Science, Office of Basic Energy Sciences Energy Frontier Research Centers program under Award Number DE-SC00211118. S.B. acknowledges the Vagelos Institute for Energy Science and Technology for the graduate fellowship. Computational support was provided by the National Energy Research Scientific Computing Center (NERSC), a U.S. Department of Energy, Office of Science User Facility located at Lawrence Berkeley National Laboratory, operated under Contract No. DE-AC02-05CH11231. 

\clearpage{} 
\bibliography{PiezoML}

\end{document}


\renewcommand{\thefigure}{S\arabic{figure}}
\renewcommand{\thetable}{S\arabic{table}}
\newpage

\section*{Methods:}

Piezoelectric calculations are performed with an energy cutoff of 680 eV, with an energy convergence of $1.4\times10^{-5}$ eV/cell and a force convergence cutoff of $2.6\times10^{-4}$ eV/\AA. Optimization steps are performed with a 4 $\times$ 4 $\times$ 4 $k$-point grid while Berry phase calculations are performed with a 4 $\times$ 4 $\times$ 7 $k$-point grid. For geometry optimization calculations in the screening phase, all parameters are kept the same, except that the energy and force conversion thresholds are raised to $1.4\times10^{-3}$ eV/cell and $2.6 \times10^{-2}$ eV/\AA, respectively, to speed up the calculations, and Gaussian smearing of 0.14 eV is added to the electronic fillings to allow for screening low to zero band-gap dopants.\par

\section{Multi-level screening protocol:}

The proposed multi-level screening protocol is designed to find dopants to improve the $e_{33}$ response of the wurtzite materials. While multiple base materials can and should be included in this search, the process is limited to dopants and combinations within one single crystal structure. The protocol for finding new materials using doping in the system of interest is as follows (Fig. \ref{fig:workflow}):
\begin{itemize}
    \item step 1: reduce a functional application to a single (or a small number) of numerical variables to optimize (here $e_{33}$ coefficient).
    \item step 2: generate a small and representative dataset (existing materials and logical extensions), then use previous work and chemical intuition to describe materials in terms of proxies that are easier to calculate than target property.
    \item step 3: use machine learning feature selection to determine which proxies to use for further screening.
    \item step 4: generate database of all possible dopant combinations and primitives for screening.
    \item step 5: identify promising candidates via proxy-based screening, and verify the predictions using more refined calculations.
    \item step 6: filter the selected candidates using thermodynamics stability analysis and practicality of the dopants. 
\end{itemize}

\section{Material Proxies}

\begin{figure}
    \centering
    \includegraphics[width=0.7\linewidth]{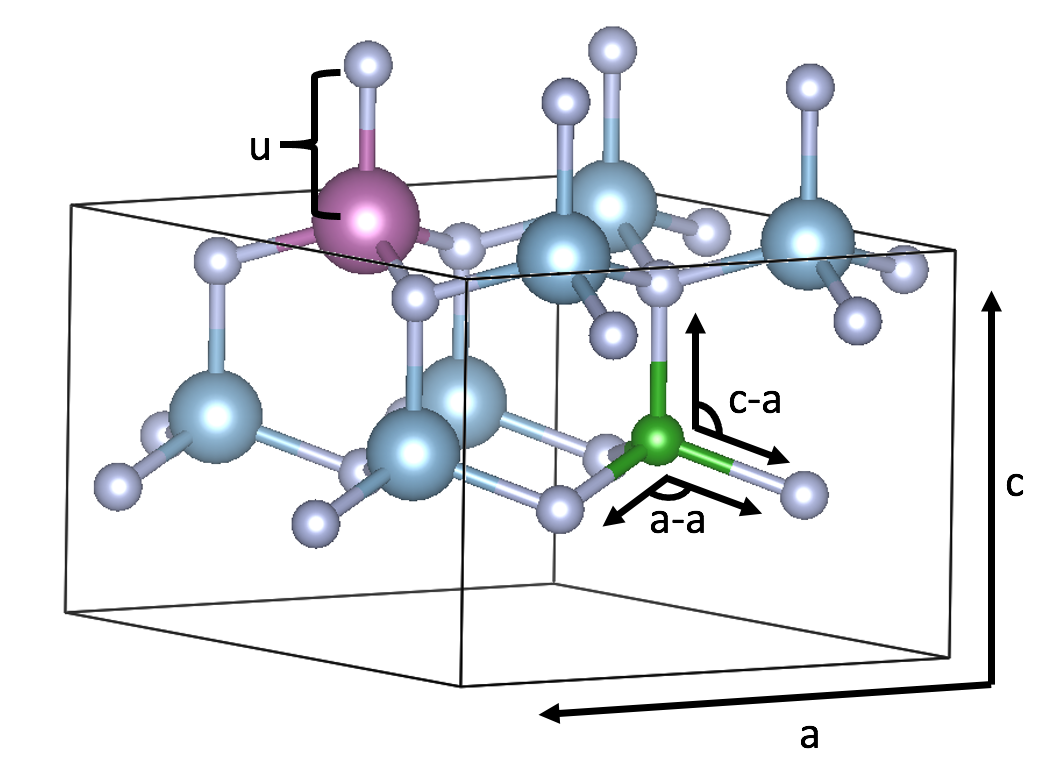}
    \caption{Example wurtzite structure with dopants. The blue atoms are the cations, replaced with the green and purple atoms as dopants, respectively, and the grey atoms are the anions.}
    \label{fig:features}
\end{figure}

The potential proxies used in level one of the screening protocol (listed in Fig. 2) represent material features obtainable through a simple optimization calculation. The lattice parameters, bond angles, and u parameters used as possible proxies are shown in Fig.\ref{fig:features}. The angles and u parameters are collected for each dopant, with the smaller atomic number dopant acting as M1. The polarization, dielectric constant, and polarizability are collected through a subsequent Berry's phase calculation on the relaxed structure.

\section{Method Testing}

In order to efficiently screen hundreds of materials, it is crucial that our computational standards are rigorous enough to be trusted, yet also as fast and computationally inexpensive as possible.

\subsection{Finite Differences Calculation for $e_{33}$}

Machine learning regression studies are particularly advantageous for maximizing a single variable at a time. Since we are interested primarily on the effect of electric field on the out-of-plane response of wutzite materials, we chose to only focus on calculating the $e_{33}$ component of the piezoelectric tenson. To calculate the entire tensor, one would have to use density functional perturbation theory. This method is valid, but requires more computational resources per calculation than our simple finite differences calculations do. Our value for $e_{33}$ of AlN using finite differences is 1.6026, compared to the reported value by the Materials Project of 1.6038, therefore, we conclude our time-saving approach is a valid method of calculating the piezoelectric tensor for wurtzites.\cite{jain2013commentary,Jong15} 

\subsection{$k$-point testing}

We increased the $k$-point grid along the c-axis in the Berry's phase calculations until the difference in calculated polarization for unstrained AlN was less than $1\times10^{-4} $ C/m$^2$, which left us with a $4\times4\times7$ grid. We found no difference within significant figures in $e_{33}$ values between using a $4\times4\times20$ and our grid, but a significant increase in computation time for the denser grid. 

\section{Datasets}

Included in the supplementary material is a csv file of all the considered proxies for the 53 materials in the initial dataset, which all are co-doped AlN (proxies.csv). Additionally, a csv of all 500 screened materials with predicted and calculated $\frac{c}{a}$ lattice ratio is included (matpredictions.csv) as well as a csv including the piezoelectric constants of the final set of materials screened (PiezoMLresults.csv). 

\section{Formation Energies}

\clearpage{} \bibliographystyle{plain}
\bibliography{PiezoML}